%% file: arxiv.tex
\documentclass{article}
\PassOptionsToPackage{numbers, compress}{natbib}

\usepackage[main, final]{neurips_2026}

\usepackage[utf8]{inputenc} 
\usepackage[T1]{fontenc}    
\usepackage{hyperref}       
\usepackage{url}            
\usepackage{booktabs}       
\usepackage{amsfonts}       
\usepackage{nicefrac}       
\usepackage{microtype}      
\usepackage[pdftex]{graphicx}
\usepackage{amsmath}
\usepackage{multirow}
\usepackage[normalem]{ulem}
\usepackage{enumitem}
\useunder{\uline}{\ul}{}
\usepackage{floatrow}
\newfloatcommand{capbtabbox}{table}[][\FBwidth]
\usepackage{tabularx}
\usepackage[most]{tcolorbox}
\usepackage{caption} 
\usepackage[table, natural]{xcolor}
\usepackage{wrapfig}

\newcommand{\benchname}{FUAG-Bench}

\title{Unified Synthesis of Compositional Speech and Sound from Free-Form Text Prompts}

\author{
Yuyue Wang$^{1}$\thanks{Contact: wangyuyue123@ruc.edu.cn}
\quad
Xihua Wang$^{1}$
\quad
Xin Cheng$^{1}$
\quad
Yijing Chen$^{1}$
\quad
\textbf{Ruihua Song}$^{1}$\thanks{Corresponding author.} \\
$^{1}$ Renmin University of China
}

\begin{document}

\maketitle

\input{sections/00-abstract}
\input{sections/01-introduction}

\input{sections/02-related_work}
\input{sections/03-method}
\input{sections/04-experiments}
\input{sections/05-conclusion}




\begin{ack}
    This work is supported by the National Natural Science Foundation of China (No.62276268) and Huawei Inc. The authors declare no competing interests.
\end{ack}

\newpage

\bibliographystyle{IEEEtran}
\bibliography{ref.bib}

\newpage
\input{sections/appendix}

\end{document}

%% file: sections/00-abstract.tex
\begin{abstract}
  Generating unified audio that naturally integrates speech and sounds from free-form text prompts remains challenging. For example, the prompt ``A fast drum roll ends with a cymbal crash, after which a man shouts: `Come on!'.'' describes several events that should appear in order and form a coherent recording rather than independent tracks that lack coherent temporal and semantic interactions. Current methods broadly follow two paradigms: pipeline methods couple separate text-to-sound and text-to-speech models, while unified models rely on a single generator that still requires structured, multi-field prompt inputs. Both paradigms require upstream prompt decomposition or rewriting, limiting free-form flexibility and risking cascading errors. To enable direct generation from free-form prompts within a single model, we introduce PlanAudio, an end-to-end framework that maps unconstrained free-form text to unified audio without intermediate prompt preprocessing. To address the challenges of direct free-form generation, PlanAudio integrates latent planning into a single autoregressive framework, predicting continuous audio layout before acoustic token synthesis. We construct a large-scale training corpus and introduce Free-Form Unified Audio Generation Benchmark (\benchname) for evaluation, both spanning speech, sound, and joint speech-and-sound generation. Experiments show that a single PlanAudio outperforms existing pipeline and unified baselines on speech-and-sound generation, while remaining competitive with specialist models on speech and sound generation. Further analysis validates the effectiveness of audio layout planning over textual planning, acoustic planning, and direct generation. Sample Available: \url{https://planaudio.github.io/PlanAudio2026/}.
\end{abstract}

%% file: sections/01-introduction.tex
\section{Introduction}
\label{sec:intro}
In real-world scenarios, speech and environmental sounds often coexist and interact within the same audio scene. While recent Text-to-Speech (TTS) \cite{cosyvoice2,cosyvoice3,Qwen3-TTS} and Text-to-Sound (T2S) \cite{AudioGen,audioldm2,stableaudio} models excel in their respective domains, they are often limited to their target task and struggle with open-domain scenarios where speech and sounds are naturally described and composed together. 

\begin{figure}
    \centering
    \includegraphics[width=0.9\linewidth]{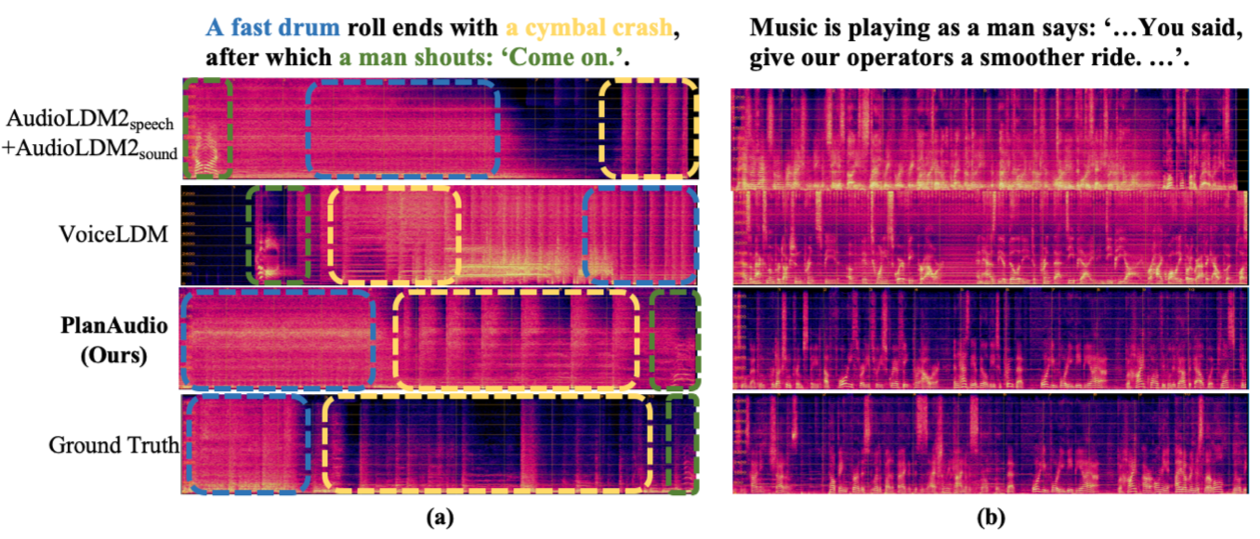}
    \caption{Qualitative comparison of audio mel-spectrograms. AudioLDM2\textsubscript{Speech} + AudioLDM2\textsubscript{Sound} denotes a pipeline approach with separate sound and speech generation, while VoiceLDM denotes a unified baseline. Textual keywords and corresponding audio regions are color-coded. (a) Pipeline and baseline models exhibit temporal misalignments (e.g., out-of-order events), whereas PlanAudio more faithfully reproduces the event sequence. (b) PlanAudio achieves more consistent relative acoustic prominence between speech and sound, closely matching the Ground Truth.}
    \label{fig:case}
\end{figure}

For instance, a free-form prompt such as ``A fast drum roll ends with a cymbal crash, after which a man shouts: `Come on!'\,'' describes multiple audio events that must be synthesized as a coherent recording with accurate temporal relationships, rather than generated independently and simply mixed together. This calls for a unified generation framework that can directly synthesize speech, sound effects, and their combinations from free-form text prompts.

A naive pipeline (e.g., AudioLDM2\textsubscript{Speech} + AudioLDM2\textsubscript{Sound}~\cite{audioldm2}) rewrites the free-form prompt into separate T2S and TTS tracks, generates them independently, and then merges them. As shown in Fig.~\ref{fig:case}, it fails to capture temporal and semantic interactions, causing misaligned event timing and unbalanced relative levels. Recent unified models~\cite{voiceldm, voicedit, controlaudio, audiobox} avoid explicit track merging but still require structured multi-inputs or external LLM rewriting. For example, VoiceLDM~\cite{voiceldm} (its generated mel-spectrogram also shown in Fig.~\ref{fig:case}) rewrites the prompt into ``speech content'' and ``environment description,'' rather than inferring from the original prompt. Such rewriting restricts free-form prompting, risks error propagation, and increases system complexity.

We therefore introduce \textbf{Free-Form-Prompt-to-Unified-Audio Generation}, where a model synthesizes speech, sound, or joint speech-and-sound from a single unstructured prompt, with no structured templates, task tags, or external rewriting. This task raises three challenges: (1) eliminating the prompt decomposition that prior paradigms depend on; (2) avoiding the omission of events specified in the prompt when mapping free-form text directly to audio; and (3) addressing the scarcity of joint speech-and-sound annotations, as existing corpora \cite{libritts,audiocaps,wavcaps,speechcraft,common_voice} treat speech and sound in isolation.

To address these challenges, we propose PlanAudio, an autoregressive LLM framework that directly processes free-form prompts using the LLM's native tokenizer, without relying on external prompt rewriting or text encoders. To tackle the difficulty of direct free-form generation, PlanAudio incorporates a latent planning mechanism into the autoregressive generation process, predicting continuous audio layout to capture the global temporal and semantic relationships of audio events prior to acoustic token synthesis. We further construct a large-scale training corpus and introduce Free-Form Unified Audio Generation Benchmark (\benchname) for evaluation, both spanning speech, sound, and joint speech-and-sound generation. \benchname \ incorporates $4500$ real-world speech-and-sound recordings with automatically generated and human-verified joint annotations. A single PlanAudio model outperforms both pipeline and unified baselines on speech-and-sound joint generation, while remaining competitive with single-task specialists on speech generation and sound generation. Further analysis confirms that our continuous audio layout plan outperforms textual planning, acoustic planning, and direct generation. Audio samples are available at \url{https://planaudio.github.io/PlanAudio2026/}.

%% file: sections/02-related_work.tex
\section{Related works}
\label{sec:related_work}
\subsection{Audio Generation: From Specific to Unified}
Text-to-Speech (TTS)~\cite{cosyvoice2,cosyvoice3,Qwen3-TTS} and Text-to-Sound (T2S)~\cite{audioldm2,AudioGen,stableaudio} focus on linguistic and environmental audio, respectively. 
Recent efforts enhance generation control via natural language, such as structuring prompts to guide temporal order in T2S~\cite{audiocomposer,freeaudio,picoaudio} or specifying instructions to control prosody and style in TTS~\cite{VoxInstruct,flexivoice,MoE-TTS}. However, these scenario-specific methods cannot model intricate interactions between speech and sounds, failing to generate unified audio from free-form text. 

To overcome this, unified models have emerged, broadly falling into two paradigms: (1) Task-level unification~\cite{uniaudio,fugatto}, which integrates multiple audio tasks into a single model but lacks speech-and-sound joint generation; and (2) Generation-level unification~\cite{voiceldm,audiobox,voicedit,controlaudio}, which can synthesize joint speech-and-sound content. However, generation-level unification always requires explicit prompt decomposition or structured templates, shifting the burden of input structuring to users or external LLMs. In contrast, PlanAudio offers a unified framework that seamlessly generates speech, sound, and speech-and-sound content from free-form text prompts without explicit prompt rewriting.

\subsection{Planning in Audio Generation}
Intermediate planning has been explored in generative models to improve the controllability and quality of complex outputs~\cite{CoT-VTM,OV-InstructTTS,XS-CoT}. In audio generation, existing methods mainly represent intermediate plans as textual reasoning chains or structured instructions. For example, CoT-VTM~\cite{CoT-VTM} generates explicit reasoning chains for vision-to-music generation, and OV-InstructTTS~\cite{OV-InstructTTS} leverages structured intermediate instructions for controllable speech synthesis. Similarly, prompt decomposition or rewriting in existing unified audio generation frameworks~\cite{voicedit,voiceldm,audiobox,freeaudio} can also be viewed as a form of explicit textual plan. They use manual or external LLMs to decompose the prompt into environment description and speech content, then feed them separately into the model. However, constructing explicit textual plans incurs additional annotation or rewriting costs, making them expensive to obtain at scale.

Instead of relying on explicit textual plans, an alternative is to represent intermediate plans in latent spaces. In particular, latent reasoning~\cite{latent_cot_llm1,latent_cot_llm2,latent_cot_survey} operates directly on continuous representations, preserving fine-grained structural information that are difficult to verbalize. This capability is especially critical for complex audio generation, where simple text cannot specify the subtle interplay between speech and sound events, such as relative loudness balance and temporal intertwining. Despite this potential, leveraging continuous semantic latents as an intermediate global plan for unified audio generation from free-form prompts remains underexplored.

%% file: sections/03-method.tex
\begin{figure}[H]
    \centering
    \includegraphics[width=0.9\linewidth]{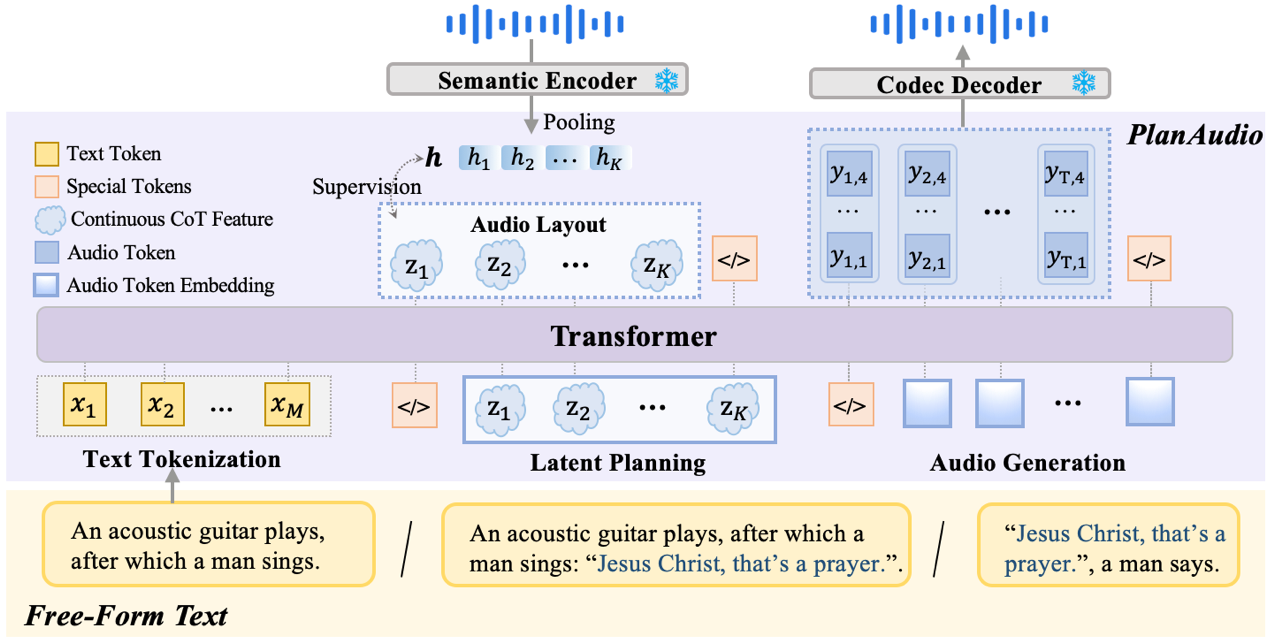}
    \caption{PlanAudio framework. Given a free-form text prompt, PlanAudio first performs latent planning to predict an audio layout, which captures the temporal and semantic organization of the target audio. The predicted audio layout then guides the autoregressive generation of hierarchical audio tokens, which are subsequently decoded into the final unified audio waveform.}
    \label{fig:model}
\end{figure}

\section{Methodology} \label{sec:method}
We formulate Free-Form-Prompt-to-Unified-Audio Generation as generating a unified audio waveform $y$ directly from a single unconstrained text prompt $x$, without external prompt decomposition or rewriting. The prompt may describe sound events, speech, or the both, and the model generates corresponding audio while preserving the specified semantic and temporal relationships.

We present \textbf{PlanAudio} (Fig.~\ref{fig:model}), an LLM-based framework that directly maps free-form text to unified audio within a single autoregressive model. Instead of explicitly decomposing the prompt into separate speech and sound instructions, we introduce a latent planning mechanism. PlanAudio first predicts a continuous audio layout $\mathbf{z}$ that captures the temporal and semantic organization of the target audio, which then guides subsequent acoustic token synthesis.

\subsection{Text and Audio Representations}
We first introduce the representations and notions used by PlanAudio:

\begin{itemize}[leftmargin=1em]
    \item The free-form text prompt $x$ is processed by the built-in LLM tokenizer without external text encoders. The resulting tokens $\mathbf{x} = \{x_1, \dots, x_M\}$ serve as the conditional prefix. 

    \item We use two representations of the target audio $y$. For \textbf{audio layout}, we extract a sequence of continuous audio features using the pre-trained Audio Flamingo 3 Encoder (AF3Encoder)~\cite{af3} and apply segment-wise mean pooling to obtain a fixed-length sequence $\mathbf{h} = \{h_1, \dots, h_K\}$, where each $h_k$ captures \textit{semantic} information from a corresponding temporal segment of the target audio. These features provide the training targets for learning the continuous audio layout $\mathbf{z}$ and are not involved in inference. For \textbf{acoustic representation}, the target audio is discretized into hierarchical tokens $\mathbf{y} = [y_1, \dots, y_N] \in \mathbb{N}^{N \times Q}$ using the AudioCraft tokenizer\footnote{https://github.com/facebookresearch/audiocraft}, where $y_n$ denotes the discrete audio tokens at time step $n$ and $Q=4$ is the number of codebooks. These tokens provide the training targets for acoustic generation.
\end{itemize}

\subsection{PlanAudio Framework} 
As illustrated in Fig.~\ref{fig:model}, PlanAudio factorizes the Free-Form-Text-Prompt-to-Unified-Audio generation into a two-phase process mediated by a continuous audio layout $\mathbf{z} = \{z_1, \dots, z_K\}$. The generation process is formulated as:

\begin{equation}
    P(\mathbf{y}, \mathbf{z} \mid \mathbf{x}) = \underbrace{P(\mathbf{z} \mid \mathbf{x})}_{\text{Latent Planning}} \cdot \underbrace{P(\mathbf{y} \mid \mathbf{x}, \mathbf{z})}_{\text{Audio Generation}},
\end{equation} 
where $\mathbf{x}$ represents the text tokens and $\mathbf{y}$ denotes the hierarchical discrete audio tokens.

To enable end-to-end processing, we organize all modalities into a single sequence, using special tokens to demarcate different segments. This allows latent planning and acoustic generation to be performed within a single Transformer backbone. Specifically, a data point is formatted as:
\begin{equation}
    \mathbf{S} = [\texttt{<|sot|>}, \mathbf{x}, \texttt{<|sol|>}, \mathbf{z}, \texttt{<|soa|>}, \mathbf{y}, \texttt{<|eoa|>}],
\end{equation}
where $\mathbf{x}$, $\mathbf{z}$, and $\mathbf{y}$ represent text tokens, continuous audio layout, and discrete audio tokens, respectively. Special tokens $\texttt{<|sot|>}$ (start of text), $\texttt{<|sol|>}$ (start of latent), $\texttt{<|soa|>}$ (start of audio), and $\texttt{<|eoa|>}$ (end of audio) serve as functional delimiters to guide the model through different phases.

\paragraph{Latent Planing.} PlanAudio predicts a continuous audio layout $\mathbf{z}$ from the textual prefix $\mathbf{x}$. 
\begin{equation}
P(\mathbf{z} \mid \mathbf{x}) = \prod_{k=1}^{K} P(z_k \mid z_{<k}, \mathbf{x}),
\end{equation}
where $K$ is a predefined constant representing the planning horizon. The predited audio layout provides a compact representation of the temporal and semantic organization of the target audio, including event orchestration and the prosodic orientation of speech. By introducing this latent planning stage, the model first establishes the global organization of the target audio before generating low-level acoustic details.

\paragraph{Acoustic Generation.} In this phase, the model generates audio tokens conditioned on both the original prompt $\mathbf{x}$ and the continuous audio layout $\mathbf{z}$ predicted in the preceding phase. Following the hierarchical structure of the discrete tokenizer, the model predicts audio tokens across $Q$ codebooks:
\begin{equation}
P(\mathbf{y} \mid \mathbf{x}, \mathbf{z}) = \prod_{n=1}^{N} \prod_{q=1}^{Q} P(y_{n,q} \mid y_{<n, \ast}, y_{n, <q}, \mathbf{x}, \mathbf{z}).
\end{equation}
Here, $\mathbf{z}$ provides global guidance for acoustic generation by conveying the intended temporal and semantic organization of the target audio, while $\mathbf{x}$ preserves direct access to the original prompt.

\subsection{Training and Inference}
\paragraph{Training Objectives.} To optimize the two-phase generation process, we introduce a dual-objective loss that supervises both latent planning and acoustic generation.


For \textbf{latent planning}, each planning step maps the LLM hidden state into the AF3 feature space by $\phi(\cdot)$, producing one frame of the latent audio layout $\mathbf{z}_{1:T}$. During training, $\psi(\cdot)$ maps the ground-truth AF3Encoder sequence $\mathbf{h}$ into the LLM input space, and the sequence is fed as the teacher-forced input with a one-frame shift. Both $\phi(\cdot)$ and $\psi(\cdot)$ are a linear layer followed by layer normalization. We train $z_t$ against $h_t$ with mean squared error (MSE) and cosine similarity:

\begin{equation}
\mathcal{L}_{\text{latent}}
=
\frac{1}{T}\sum_{t=1}^{T}
\left[
\|z_t - h_t\|_2^2
+ \lambda \left(
1 - \frac{\langle z_t, h_t \rangle}{\|z_t\|\,\|h_t\|}
\right)
\right],
\end{equation}
where MSE ensures Euclidean proximity while cosine similarity encourages directional alignment.

For \textbf{audio generation}, the $Q$ codebooks at each step are predicted in parallel. Cross-entropy loss maximizes the likelihood of the target audio tokens given the text $\mathbf{x}$, the teacher-forced layout $\mathbf{h}$, and the tokens from previous steps:
\begin{equation}
\mathcal{L}_{\text{audio}} = - \sum_{n=1}^{N} \sum_{q=1}^{Q} \log P(y_{n,q} \mid y_{<n}, \mathbf{x}, \mathbf{h}).
\end{equation}
The final training objective is a weighted sum of the layout and acoustic losses: $\mathcal{L}_{\text{total}} = \lambda_1 \mathcal{L}_{\text{latent}} + \lambda_2 \mathcal{L}_{\text{audio}}$, where $\lambda_1$ and $\lambda_2$ are hyperparameters balancing semantic planning and generation quality.

\paragraph{Inference.}
Inference proceeds in two phases. Given the free-form text prompt $\mathbf{x}$, PlanAudio autoregressively predicts the continuous audio layout $\mathbf{z}$ for $K$ steps. At each step, $\phi(\cdot)$ maps the LLM hidden state to a layout frame $z_k$, and $\psi(z_k)$ maps that frame into the LLM input space as the input for the next planning step, using the same input mapping as teacher forcing during training. After the planning phase, the \texttt{<|soa|>} token triggers acoustic generation. Conditioned on both $\mathbf{x}$ and $\mathbf{z}$, PlanAudio autoregressively synthesizes audio token sequence $\mathbf{y}$ until the \texttt{<|eoa|>} token is generated, after which the tokens are decoded into the final waveform.

%% file: sections/04-experiments.tex
\section{Experiments} \label{sec:exp}

\subsection{Implementation Details}
PlanAudio is initialized from Qwen2.5-1.5B\footnote{\url{https://huggingface.co/Qwen/Qwen2.5-1.5B}}, which provides the text tokenizer and autoregressive backbone, with no external text encoder or rewriter. Target audio is discretized by the AudioCraft tokenizer ($4$ codebooks, $2048$ entries, $50$ Hz), and the resulting tokens are predicted under the delayed token interleaving pattern~\cite{musicgen} (App.~\ref{app:delay}). Semantic supervision comes from AF3Encoder~\cite{af3}, whose $750$ output frames are mean-pooled every $125$ frames into $K=6$ anchors, and the planning and acoustic losses are weighted equally. The planning horizon $K$ is fixed, while the audio length is determined by when \texttt{<|eoa|>} is generated, rather than by a preset length.

Training samples are drawn from the three scenarios (sound, speech, speech-and-sound) in a prescribed ratio, and one epoch contains as many samples as the sum of the three datasets. We use a maximum batch size of $16$ with a token budget of $4000$ per batch, and accumulate gradients over $8$ steps. All parameters are trained with Adam at an initial learning rate of $10^{-4}$, using a $3000$-step warmup followed by inverse square-root decay with a decay coefficient of $0.1$. Training runs for $70$ epochs on $8$ NVIDIA A800 GPUs ($80$~GB) and takes about $10$ days. Inference uses top-$k$ sampling with $k=25$. We also report, in App.~\ref{app:codec_exp}, the codec reconstruction comparison behind the tokenizer choice and the trade-off with generative performance, and analyze the choices of planning horizon $K$ and the loss weights in App.~\ref{app:ablation_planning_horizon} and App.~\ref{app:ablation_loss_weight}.

\subsection{Data Construction and Benchmark}
\label{sec:data}
Free-form-prompt-to-unified-audio generation takes no task label: the model infers from the prompt alone whether speech, sound, or both are requested. For analysis only, we group prompts into three scenarios: \textbf{sound}, \textbf{speech}, and \textbf{speech-and-sound}. There is a bottleneck in joint annotation: public corpora provide either speec transcripts~\citep{libritts,common_voice} or sound descriptions~\citep{audiocaps,wavcaps}, but never both for the same recording. We therefore build our own training corpus and benchmark, using only real recordings and generating only the prompts.

\paragraph{Training corpus.}
For \textbf{speech-and-sound}, we use raw AudioSet recordings~\citep{audioset} to preserve real-world speech-sound interaction. Automatically writing free-form prompts for these recordings in a single pass is prone to hallucination in both scene and speech, so we generate the two parts separately and then merge them: Whisper-large~\citep{whisper} transcribes the speech, and Gemini-2.5 Pro describes the scene with a \texttt{<speech content>} placeholder that is then filled with the transcript. The retained \textit{description-transcript pairs} later serve as the Oracle-Split input for baselines (Sec.~\ref{sec:eval_setup}). \textbf{Sound} and \textbf{speech} data preserve single-scenario ability and teach the model to tell scenarios apart without tags. Sound prompts are rewritten from AudioCaps~\cite{audiocaps} and WavCaps~\cite{wavcaps} captions; speech prompts combine LibriTTS~\cite{libritts} and Common Voice~\cite{common_voice} transcripts with speaker attributes and an explicit mention of a quiet environment. We also add speech segments extracted from the speech-and-sound recordings, so that identical transcripts appear with and without their acoustic scene and only the prompt can distinguish them. Since free-form prompts can describe the same audio in many ways, each clip is paired with $5$ prompt variants to improve flexibility and generalization to unseen phrasings. The corpus contains $1.43$M audio clips ($372$k speech-and-sound, $451$k sound, $605$k speech). Please refer to App.~\ref{app:data_prompts} for the specific data construction prompt.

\paragraph{Evaluation benchmark.}
No existing benchmark evaluates unified audio generation from free-form prompts. We therefore introduce the \textbf{Free-Form Unified Audio Generation Benchmark (\benchname)}, in which each test item pairs a free-form prompt with a real recording. Its three subsets follow the scenarios above. The \textbf{Sound} subset uses the AudioCaps~\cite{audiocaps} test set with its original captions. The \textbf{Speech} subset uses the LibriTTS~\cite{libritts} test set, with prompts generated in the same way as for the training data and one of the $5$ variants kept at random per clip. The \textbf{Speech-and-sound} subset uses $4500$ held-out AudioSet~\cite{audioset} clips, with prompts generated and sampled likewise; these clips and their derived speech segments are excluded from all training data. For the speech-and-sound subset, we further provide a human audit of annotation quality in App.~\ref{app:data_annotation_quality}, and the distribution of speech-sound interaction types in App.~\ref{app:data_interaction_types}.

\subsection{Evaluation Setup} \label{sec:eval_setup}

\paragraph{Metrics.}
We evaluate a single PlanAudio checkpoint on all three subsets of \benchname{} (Sec.~\ref{sec:data}), using the same free-form input throughout. Objective metrics mainly cover two dimensions. (1) Speech-related (Speech \& Speech-and-Sound): Word Error Rate (WER) for intelligibility, and UTMOS for perceived voice quality. (2) Sound-related (Sound \& Speech-and-Sound): Fr\'echet distance (FD) and KL divergence, each computed with PANNs~\cite{panns} and PaSST~\cite{passt}, for audio fidelity; Inception Score (IS) for diversity; and CLAP~\cite{clap} similarity for text-audio alignment. To evaluate temporal structure of sound events, we follow \citep{tiva} and report the Wasserstein distance (W-dis) between the onset detection function curves of the generated and reference audio, where lower values indicate closer timing. To measure whether sound events named in the prompt are missing from the audio, we also report the Semantic Coverage Factor (SCF; see details in App.~\ref{app:eval_scf}).

Since automatic metrics cannot fully judge whether the audio follows the prompt or whether speech and sound blend naturally, we further conduct a subjective test on 50 random speech-and-sound samples. Ten listeners rate each sample from 0 to 5 on \textit{Acoustic Quality}, \textit{Temporal Correctness}, \textit{Semantic Alignment}, and \textit{Authenticity}.

\paragraph{Baselines.}
No prior model accepts free-form prompts, so we compare with three types of models, each using its own input format. (1) \textit{Unified}: VoiceLDM~\cite{voiceldm}, which takes a transcript and an environment description and can generate all three scenarios\footnote{Audiobox~\cite{audiobox}, VoiceDit~\cite{voicedit}, and ControlAudio~\cite{controlaudio} also accept structured inputs but lack open-source checkpoints, restricting our comparisons to available models.}. (2) \textit{Specialists}: AudioLDM2~\cite{audioldm2}, Tango~\cite{tango}, and Make-An-Audio~\cite{make-an-audio} for sound; PromptTTS++~\cite{prompttts++} and CosyVoice2~\cite{cosyvoice2} for speech. (3) \textit{Pipelines}: for speech-and-sound, AudioLDM2\textsubscript{Speech} + AudioLDM2\textsubscript{Sound} (AudioLDM2\textsubscript{S+Sp}), whose two outputs are mixed.

For sound subset and speech subset, specialists receive their native inputs; VoiceLDM receives the scene description or transcript in the matching condition, with the other left empty. For the speech-and-sound subset, every baseline needs the prompt split into a scene description and a transcript. Because the quality of this split directly affects the generated audio, we obtain it in three ways: with Qwen3-0.6B~\cite{qwen3} as a lightweight local rewriter, with Gemini-2.5 Pro as a strong API-based rewriter, and with Oracle-Split, the \textit{description-transcript pairs} kept during data construction (Sec.~\ref{sec:data}). The setting used by each baseline is listed in the result tables.

\subsection{Main Results and Analysis}
All results below come from the same PlanAudio checkpoint.  

\begin{table}
\caption{Results on the speech-and-sound subset of \benchname. Sound- and Speech-related metrics are evaluated using official checkpoints. Best results are in \textbf{bold}; second-best are \underline{underlined}. PlanAudio achieves superior overall performance, dominating sound generation while remaining competitive in speech.
}
\label{tab:main-composite}
\centering
\resizebox{\linewidth}{!}{
\begin{tabular}{lccccccccc}
\toprule
\multirow{2}{*}{Model} & \multirow{2}{*}{Decomposition} & \multicolumn{6}{c}{Sound-related} & \multicolumn{2}{c}{Speech-related} \\ \cmidrule(lr){3-8} \cmidrule(lr){9-10}
& & \multicolumn{1}{l}{FD\textsubscript{PANNs} $\downarrow$} & \multicolumn{1}{l}{FD\textsubscript{PaSST} $\downarrow$} & \multicolumn{1}{l}{KL\textsubscript{PaSST} $\downarrow$} & \multicolumn{1}{l}{CLAP $\uparrow$} & \multicolumn{1}{l}{SCF $\uparrow$} & \multicolumn{1}{l}{W-dis $\downarrow$} & \multicolumn{1}{c}{WER $\downarrow$} & \multicolumn{1}{c}{UTMOS $\uparrow$} \\ \midrule
GroundTruth & - & 0.00 & 0.00 & 0.00 & 0.17 & 0.50 & 0.0 & 0.10 & 2.69 \\
Reconstruction & - & 3.82 & 112 & 0.23 & 0.21 & - & - & 0.21 & 2.49 \\ \midrule
AudioLDM2\textsubscript{S+Sp}~\cite{audioldm2} & Oracle-Split & \underline{14.3} & \underline{240} & \underline{1.10} & \underline{0.21} & 0.43 & 3.97 & 0.71 & 2.16 \\ 
VoiceLDM-s~\cite{voiceldm} & Oracle-Split & 25.2 & 379 & 1.39 & 0.15 & 0.25 & 5.38 & 0.70 & 2.35 \\ 
VoiceLDM-m~\cite{voiceldm} & Oracle-Split & 23.4 & 343 & 1.34 & 0.19 & 0.38 & 5.14 & \textbf{0.13} & \textbf{2.81} \\ \midrule
\multirow{2}{*}{VoiceLDM-m\cite{voiceldm}} & Gemini-2.5-pro & 23.7 & 347 & 1.34 & 0.19 & 0.35 & 5.21 & \underline{0.19} & 2.71 \\
 & Qwen3-0.6B & 23.8 & 362 & 1.36 & 0.15 & 0.30 & 5.26 & 0.27 & \underline{2.73} \\ \midrule
PlanAudio & None & \textbf{8.96} & \textbf{197} & \textbf{0.98} & \textbf{0.22} & \textbf{0.45} & \textbf{3.04} & 0.30 & 2.53 \\ \bottomrule
\end{tabular}
}
\end{table}

\begin{table}
    \caption{Subjective results for speech-and-sound subset of \benchname. Ratings (0-5) across four criteria: Acoustic Quality (Quality), Temporal Correctness (Temporal), Semantic Alignment (Semantic), and Authenticity. Confidence intervals are 95\%. PlanAudio performs best in all criteria.}
    \label{tab:user_study}
    \centering
    \resizebox{\linewidth}{!}{\begin{tabular}[b]{lccccc}
        \toprule
        Model & Decompisition & Quality $\uparrow$ & Temporal $\uparrow$ & Semantic $\uparrow$ & Authenticity $\uparrow$ \\ \midrule
        VoiceLDM-s~\cite{voiceldm} & Oracle-Split & 2.78 $\pm$ 0.30 & 2.65 $\pm$ 0.24 & 2.67 $\pm$ 0.25 & 2.71 $\pm$ 0.28 \\ 
        VoiceLDM-m~\cite{voiceldm} & Oracle-Split & 2.83 $\pm$ 0.11 & 2.78 $\pm$ 0.13 & 2.95 $\pm$ 0.13 & 2.93 $\pm$ 0.17 \\ 
        AudioLDM2\textsubscript{S+Sp}~\cite{audioldm2} & Oracle-Split  & 2.24 $\pm$ 0.31 & 2.20 $\pm$ 0.32 & 2.38 $\pm$ 0.36 & 2.40 $\pm$ 0.39 \\ \midrule
        PlanAudio & None & \textbf{3.23} $\pm$ 0.13 & \textbf{3.16} $\pm$ 0.16 & \textbf{3.36} $\pm$ 0.13 & \textbf{3.47} $\pm$ 0.21 \\ \bottomrule  
        \end{tabular}}
\end{table}

\begin{table}
\caption{ 
Results on the sound subset of \benchname, evaluated on Sound-related dimensions only.
Results are derived from official checkpoints, with the highest scores in \textbf{bold} and the second highest \underline{underlined}. \colorbox[HTML]{D3D3D3}{Rows in gray} denote specialist baselines designed solely for text-to-sound. PlanAudio surpasses unified baselines on all metrics and maintains competitive 
against specialist models.}
\label{tab:main-sound}
\centering
\resizebox{1.0\linewidth}{!}{
\begin{tabular}{lcccccccc}
\toprule
\multicolumn{1}{c}{Model} & \multicolumn{1}{l}{FD\textsubscript{PANNs} $\downarrow$} & \multicolumn{1}{l}{FD\textsubscript{PaSST} $\downarrow$} & \multicolumn{1}{l}{KL\textsubscript{PANNs} $\downarrow$} & \multicolumn{1}{l}{KL\textsubscript{PaSST} $\downarrow$} & \multicolumn{1}{l}{IS $\uparrow$} & \multicolumn{1}{l}{CLAP $\uparrow$} & \multicolumn{1}{l}{SCF $\uparrow$} & \multicolumn{1}{l}{W-dis $\downarrow$} \\ \midrule
GroundTruth & 1.05 & 0.29 & 0.02 & 0.00 & 13.2 & 0.29 & 0.35 & 0.0 \\
Reconstruction & 11.4 & 130 & 0.49 & 0.48 & 9.11 & 0.23 & - & - \\  \midrule
\rowcolor[HTML]{D3D3D3}[7pt][7pt]  AudioLDM2~\cite{audioldm2} & 32.5 & 395 & \underline{1.56} & \underline{1.51} & \textbf{8.54} & \textbf{0.21} & 0.29 & 4.30  \\
\rowcolor[HTML]{D3D3D3}[7pt][7pt]
Make-An-Audio~\cite{make-an-audio} & 27.9 & \textbf{182} & 1.60 & 1.62 & 7.44 & \textbf{0.21} & 0.27 & 4.13 \\
\rowcolor[HTML]{D3D3D3}[7pt][7pt] 
Tango~\cite{tango} & \underline{26.1} & 276 & \textbf{1.37} & \textbf{1.29} & \underline{8.23} & \underline{0.19} & \textbf{0.33} & \textbf{3.78} \\ 
VoiceLDM-s~\cite{voiceldm} & 58.4 & 430 & 3.27 & 3.01 & 4.41 & 0.10 & 0.14 & 6.23 \\
VoiceLDM-m~\cite{voiceldm} & 55.8 & 433 & 3.37 & 3.05 & 4.18 & 0.07 & 0.15 & 4.88 \\ \midrule
PlanAudio & \textbf{24.7} & \underline{233} & 1.93 & 1.89 & 8.02 & \underline{0.19} & \underline{0.32} & \underline{3.89} \\ \bottomrule
\end{tabular}}
\end{table}

\paragraph{Superiority in Speech-and-Sound Scenario.}
Tab.~\ref{tab:main-composite} and Tab.~\ref{tab:user_study} report objective and subjective results on the speech-and-sound subset. This subset tests the core difficulty of the task: the model must realize every prompted event and place speech and sound in the right order and balance. 

On sound-related metrics, PlanAudio achieves the best FD and KL among all baselines, even when they receive Oracle-Split. It also achieves the lowest W-dis ($3.04$ vs.\ $3.97$ for the best baseline), indicating event timing closest to the reference, and the highest SCF ($0.45$, close to $0.50$ for the ground truth), indicating that it omits the fewest prompted events. The subjective test agrees: PlanAudio scores highest on all four criteria, with clear margins in temporal correctness and semantic alignment (Tab.~\ref{tab:user_study}). Fig.~\ref{fig:case} also illustrates both aspects. In case (a), PlanAudio follows the prompted event order, whereas the baselines place events out of order. In case (b), PlanAudio matches the relative loudness of speech and sound in the ground truth, whereas the baselines do not.

\begin{wraptable}{r}{0.4\linewidth} 
  \centering
  \caption{Results on the speech subset of \benchname, evaluated on Speech-related metrics only. All results are derived from official checkpoints, with the highest scores in \textbf{bold} and the second highest \underline{underlined}. \colorbox[HTML]{D3D3D3}{Rows in gray} denote specialist baselines designed solely for text-to-speech. PlanAudio outperforms unified baselines on all metrics and maintains competitive performance against specialist baselines.}
    \label{tab:main_speech}
    \resizebox{1.0\linewidth}{!}{\begin{tabular}[b]{lcc}
    \toprule
    Model & WER $\downarrow$ & UTMOS $\uparrow$ \\ \midrule
    GroundTruth & 0.03 & 3.69 \\
    Reconstruction & 0.04 & 3.13 \\ \midrule
    \rowcolor[HTML]{D3D3D3}[7pt][7pt] 
    Prompt TTS++~\cite{prompttts++} & 0.12 & \textbf{3.51} \\ 
    VoiceLDM-s~\cite{voiceldm} & 0.62 & 2.75 \\
    VoiceLDM-m~\cite{voiceldm} & 0.13 & 2.99 \\ \midrule
    PlanAudio & \underline{0.11} & \underline{3.11} \\ \bottomrule
    \end{tabular} }
\end{wraptable}
On speech-related metrics, PlanAudio falls behind VoiceLDM-m, mainly for two reasons. First, the codec limits speech quality: reconstructed audio already has a WER of $0.21$. Second, VoiceLDM's training data include clean speech overlaid onto separately recorded sound, which keeps speech easy to recognize but makes the scene less natural, whereas PlanAudio is trained on real recordings, where speech is mixed with the scene. This trade-off appears in the subjective \textit{Authenticity} score, where PlanAudio leads VoiceLDM-m by $0.54$. Moreover, VoiceLDM-m's intelligibility depends on the quality of the prompt split: when the Oracle-Split is replaced by Gemini-2.5 Pro and Qwen3-0.6B, its WER rises from $0.13$ to $0.19$ and $0.27$, whereas PlanAudio reads raw free-form prompts.

\paragraph{Generalization to Domain-Specific Audio Synthesis.}
Tab.~\ref{tab:main-sound} and Tab.~\ref{tab:main_speech} report performance on the sound subset and the speech subset. They evaluate whether unifying all three scenarios in a single model degrades quality when the prompt requests only sound or only speech. Across both subsets, PlanAudio consistently outperforms the unified baseline VoiceLDM and remains competitive with task-specific specialists, so unification does not degrade quality. On the sound subset, PlanAudio achieves CLAP, SCF, and W-dis scores close to those of the best specialist, showing that it captures both the events in the prompt and their timing. On the speech subset, PlanAudio surpasses VoiceLDM on both WER and UTMOS, unlike on the speech-and-sound subset, highlighting its high-fidelity synthesis when the prompt requests only speech. Since PlanAudio takes free-form prompts without task tags, results also show that it infers from the prompt alone whether sound or speech is requested.

\subsection{Does the Latent Plan Really Matter?}
\label{sec:oracle_cot} 
To test whether the latent planning mechanism is doing real work, we apply two inference interventions to the fully trained model. \textbf{GT-Plan} replaces the predicted audio layout $\mathbf{z}$ with the AF3 feature $\mathbf{h}$ of the ground-truth audio, removing planning error; \textbf{Reverse-Plan} supplies the AF3 feature $\mathbf{h}$ in reversed order, preserving content while destroying temporal structure. Results are shown in Tab.~\ref{tab:upper_bound}. 

Compared with PlanAudio, GT-Plan substantially reduces FD, from $197$ to $140$ for Speech-and-Sound and from $233$ to $176$ for Sound, closing approximately $67\%$ and $55\%$ of the respective gaps to the Reconstruction reference. This indicates that more accurate latent plans substantially improve distributional fidelity. However, the improvement in intelligibility is relatively limited (Speech-and-Sound WER: $0.30$ $\to$ $0.28$, compared with $0.21$ for Reconstruction), suggesting that planning accuracy alone is insufficient to fully recover intelligibility. This residual gap may also reflect limitations in the phonetic information retained by the latent representation and errors in the generation process. Conversely, Reverse-Plan substantially degrades performance across all three scenarios. This highlights the importance of temporal ordering in the latent plan: although PlanAudio has no explicit timing module, the ordering of latent features implicitly organizes audio events over time. Disrupting this order while preserving the feature content severely compromises generation quality. Together, these interventions demonstrate that the latent plan encodes not only event-level content but also its temporal organization.

\begin{table}[H]
\caption{Analysis of the latent planning mechanism. \textit{GT-Plan} replaces the predicted audio layout $\mathbf{z}$ with AF3 feature $\mathbf{h}$ extracted from the ground-truth audio; \textit{Reverse-Plan} supplies the same $\mathbf{h}$ in reversed temporal order. Both are inference interventions on the fully trained model. FD and KL are computed via the PaSST classifier.}
\label{tab:upper_bound}
\resizebox{\linewidth}{!}{\begin{tabular}{lcccccccccc}
\toprule
\multicolumn{1}{c}{} & \multicolumn{5}{c}{Speech-and-Sound} & \multicolumn{3}{c}{Sound} & \multicolumn{2}{c}{Speech} \\ \cmidrule(lr){2-6} \cmidrule(lr){7-9} \cmidrule(lr){10-11}
\multicolumn{1}{c}{\multirow{-2}{*}{Model}} & \multicolumn{1}{l}{FD $\downarrow$} & \multicolumn{1}{l}{KL $\downarrow$} & \multicolumn{1}{l}{CLAP $\uparrow$} & \multicolumn{1}{c}{WER $\downarrow$} & \multicolumn{1}{c}{UTMOS $\uparrow$} & \multicolumn{1}{l}{FD $\downarrow$} & \multicolumn{1}{l}{KL $\downarrow$} & \multicolumn{1}{l}{CLAP $\uparrow$} & \multicolumn{1}{l}{WER $\downarrow$} & \multicolumn{1}{l}{UTMOS $\uparrow$} \\  \midrule
\rowcolor[HTML]{D3D3D3}[7pt][7pt] Reconstruction & 112 & 0.37 & 0.21 & 0.21 & 2.49 & 130 & 0.48 & 0.23 & 0.04 & 3.13 \\ 
\rowcolor[HTML]{CFE2F3}[7pt][7pt]  PlanAudio & 197 & 0.98 & 0.22 & 0.30 & 2.53 & 233 & 1.89 & 0.19 & 0.11 & 3.11 \\
GT-Plan & 140 & 0.61 & 0.22 & 0.28 & 2.51 & 176 & 1.24 & 0.21 & 0.09 & 3.05 \\
Reverse-Plan & 235 & 1.03 & 0.20 & 0.67 & 2.45 & 288 & 2.03 & 0.16 & 0.77 & 3.13 \\ \bottomrule
\end{tabular}}
\end{table}

\subsection{Analysis of Latent Planning Mechanism Design}
We compare five different Planning mechanisms: (1) \textit{PlanAudio}: latent planning in continuous semantic space; (2) \textit{w/o Planning}: direct audio generation from text without an intermediate planning; (3) \textit{Textual Planning}: a natural language reasoning chain (generated by Gemini-2.5 Pro) serving as a textual bridge for planning supervision; (4) \textit{Acoustic Planning}: latent planning in continuous acoustic space, using codebook-based acoustic vectors as supervision; (5) \textit{Discrete Planning}: latent planning in discrete semantic space, supervised by a semantic token sequence generated from k-means quantization of AF3 features. All variants follow the main experimental configuration, but are trained for 50 epochs on a one-third data subset. Tab.~\ref{tab:ab_cot} yields several insights:

\begin{table}[H]
\centering
\caption{Planning mechanism comparison on three subsets of \benchname. FD and KL metrics are computed via the PasST classifier. Results indicate that PlanAudio’s semantic latent planning achieves superior performance.}
\label{tab:ab_cot}
\resizebox{\linewidth}{!}{\begin{tabular}{lccccccccccc} 
\toprule
\multicolumn{1}{c}{} & \multicolumn{6}{c}{Speech-and-Sound} & \multicolumn{3}{c}{Sound} & \multicolumn{2}{c}{Speech} \\ \cmidrule(lr){2-7} \cmidrule(lr){8-10} \cmidrule(lr){11-12}
\multicolumn{1}{c}{\multirow{-2}{*}{Model}} & \multicolumn{1}{l}{FD $\downarrow$} & \multicolumn{1}{l}{KL $\downarrow$} & \multicolumn{1}{l}{CLAP $\uparrow$} & \multicolumn{1}{c}{WER $\downarrow$} & \multicolumn{1}{c}{UTMOS $\uparrow$} & \multicolumn{1}{c}{SCF $\uparrow$} & \multicolumn{1}{l}{FD $\downarrow$} & \multicolumn{1}{l}{KL $\downarrow$} & \multicolumn{1}{l}{CLAP $\uparrow$} & \multicolumn{1}{l}{WER $\downarrow$} & \multicolumn{1}{l}{UTMOS $\uparrow$} \\  \midrule
\rowcolor[HTML]{CFE2F3}[7pt][7pt] 
PlanAudio & \textbf{177} & \textbf{1.07} & \textbf{0.20} & \textbf{0.86} & \textbf{2.24} & \textbf{0.34} & \textbf{242} & \textbf{1.94} & \textbf{0.17} & \textbf{0.35} & \textbf{3.05} \\ 
-- w/o Planning & \underline{217} & 1.43 & 0.15 & \underline{0.92} & 2.16 & 0.12 & 299 & 3.26 & 0.03 & \underline{0.50} & \underline{2.90} \\
-- w/ Explicit Planning & 230 & 1.38 &  \underline{0.16} & 1.12 & 2.13 & 0.20 & \underline{279} & \underline{3.10} & \underline{0.05} & 1.29 & 2.34 \\
-- w/ Acoustic Planning & 319 & \underline{1.37} & 0.14 & 1.21 & \textbf{2.24} & 0.09 & 374 & 3.21 & 0.04 & 2.68 & 2.70 \\
-- w/ Discrete Planning & 213 & 1.32 & 0.17 & 1.19 & 2.18 & 0.12 & 254 & 3.10 & 0.05 & 1.88 & 2.33 \\
 \bottomrule
\end{tabular}}
\end{table}

(1) \textbf{Effectiveness of semantic latent planning.} \textit{PlanAudio} is consistently the best across all subset of \benchname, confirming that a semantic latent space bridges the text-audio modality gap more effectively than textual or acoustic representations.
(2) \textbf{Expressive limits of textual and discrete planning.} Although \textit{textual planning} benefits sound generation, it suffers from an expressive bottleneck when modeling dense, overlapping speech-and-sound interactions, which our latent method successfully captures. Additionally, PlanAudio outperforms \textit{discrete planning} by preserving fine-grained acoustic nuances and semantic density, whereas quantization-induced information loss disrupts acoustic continuity.
(3) \textbf{The Reconstruction-Semantic Mismatch.} \textit{Acoustic planning} performs worst overall, as codec-based features optimized for reconstruction lack the semantic depth required for planning. Unlike semantic latents, they fail to bridge the gap between textual prompts and acoustic realization.
(4) \textbf{Robustness to Semantic Omission.}
\textit{PlanAudio} achieves the highest SCF, proving that the semantic plan effectively acts as a robust semantic anchor to ensure accurate event synthesis.

%% file: sections/05-conclusion.tex
\section{Conclusion}
\label{sec:conclusion}
We present PlanAudio, a framework for unified audio generation from free-form text. The model performs latent planning in a continuous semantic space, bridging high-level semantics and low-level acoustic synthesis. We also introduce a large-scale corpus and \benchname\ for free-form-prompt-to-unified-audio traning and evaluation. PlanAudio surpasses baselines in speech-and-sound joint generation while remaining competitive in speech and sound. Future work will target stronger audio codecs and better semantic-acoustic trade-offs. We also acknowledge the risk of misuse for deceptive audio generation.

%% file: sections/appendix.tex
\appendix

\section{Technical appendices and supplementary material}
\subsection{Data Construction Prompts} \label{app:data_prompts}
We present detailed prompts tailored for our data construction pipeline, comprising three key components: (1) prompts for speech-and-sound annotation; (2) prompts for sound caption enhancement; (3) prompts for speech caption enhancement. 

\newtcolorbox{closedprompt}{
    sharp corners,           
    colback=white,           
    colframe=black,          
    boxrule=0.8pt,           
    left=1.5em,              
    right=1.5em,             
    top=1em,                 
    bottom=1em,              
    enhanced,                
}

\begin{center} \label{app:prompts}
    \vspace{-3pt} 
    \begin{closedprompt}
    \textbf{Prompts for Speech-and-Sound Annotation}
    
    The user will input an audio track. Briefly describe the audio, focusing exclusively on what can be perceived aurally, ignore those things that cannot be verified from the audio. Generate five diverse descriptions for the same audio, separate them by inserting ``\verb|\t|'' in the end.
        \begin{itemize}[leftmargin=1.5em]
            \item Describe audio events in order, but avoid specific timestamps. 
        
            \item The audio always contains speech (including singing), identify the primary speaker(s) based on the audio, and always use the placeholders ``\texttt{<speech content>}'' to represent their transcriptions. 
            
            \item Do not describe the topic of the speech, do not transcribe the speech. Do not use phrases like ``continues to speak'' or ``continues singing'', represent the entire speech content of a primary speaker with a single ``\texttt{<speech content>}'', unless there are interleaved conversations among multiple primary speakers. 
            
            \item Background voices (indistinct chatter, crowd noise) should be described as environmental noise (e.g., ``the crowd hummed'') and do not require placeholders.
            
            \item Avoid using indicative phrases like ``The audio contains''. 
    
        \end{itemize}

    \textbf{Prompts for Sound Caption Enhancement}
    The user will input an audio clip and its original caption. Please generate 5 alternative diverse audio captions. Separate different captions with ``\verb|\t|''.
    \begin{itemize}[leftmargin=1.5em]
        \item If there are multiple audio events, please describe them in order, but avoid specific timestamps.
        \item If the audio contains speech, just consider it as a sound event and there is no need to describe the content of the speech.
    \end{itemize}

    \textbf{Prompts for Speech Caption Enhancement}
    The user will provide a set of tags describing a speech recording. There is no actual audio file. Write five distinct, one-sentence descriptions for this speech solely on the tags. Separate different sentences with ``\verb|\t|''.  \
    \begin{itemize}[leftmargin=1.5em]
        \item Use neutral narrative verbs like ``says'', ``speaks'', avoid performance-based words like ``recites'' or ``delivers''. Do not use technical terms like ``energy'', ``pitch'' as nouns.
        \item Vary the sentence structures. Do not feel obligated to use every tag in every sentence. \item Always include the placeholder ``\texttt{<speech content>}'' as the spoken text, formatted like dialogue or a statement. 
        \item The setting is a quiet environment, weave this naturally into some of the sentences (e.g., ``In the silence of the room'', ``There's no background sound.'' or ``In a quiet space'').
    \end{itemize}
    \end{closedprompt}
\end{center}

\subsection{Impact of Data Curriculum}
\begin{figure}[H]
    \centering
    \includegraphics[width=0.9\linewidth]{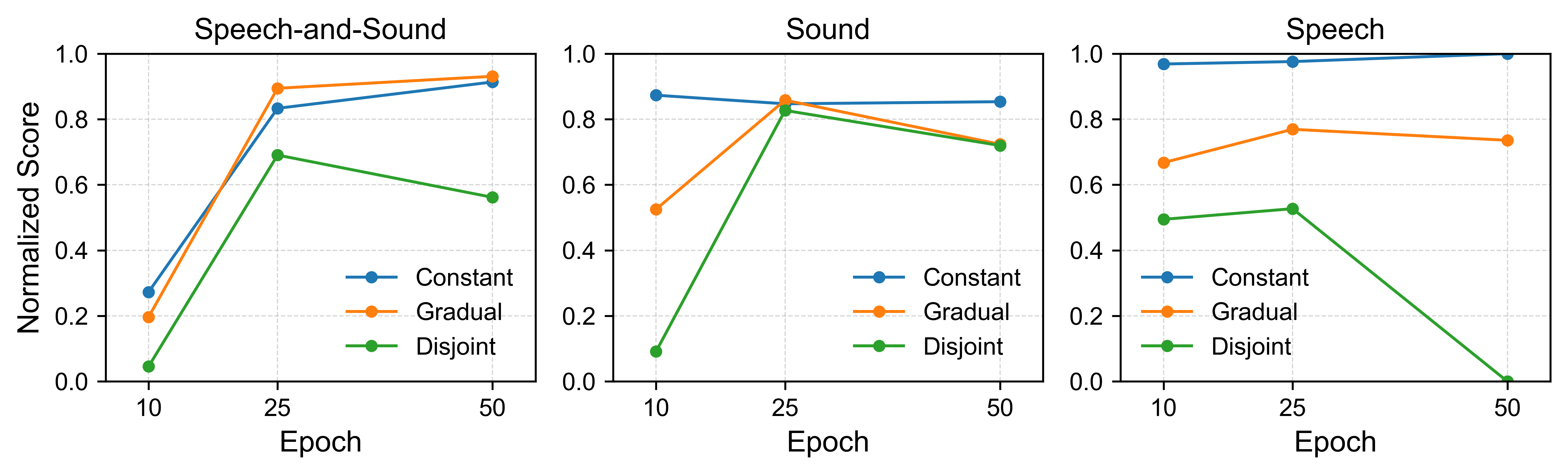}
    \caption{Normalized Score of data curriculum strategies across training epochs. Each sub-figure presents results for a scenario. This score averages strategy-wise normalized metrics; higher values indicate superior performance. Constant leads, proving uniform curricula ensure generalization.}
    \label{fig:ab_data_exp}
\end{figure}

We compare three data curriculum strategies (Tab.~\ref{tab:ab_data_schedules_detailed}) to investigate their impact on multi-scenario learning: (1) \textit{Constant}: A fixed, equal ratio of all scenarios; (2) \textit{Gradual}: Progressively increasing the proportion of speech-and-sound data; and (3) \textit{Decoupled}: Sequential training from single-scenario to composite audio. All variants follow the main experimental configuration, but are trained for 50 epochs on a one-third data subset. To facilitate cross-category comparison, we report a Normalized Score, calculated by averaging min-max normalized metrics within each scenario. A higher Normalized Score indicates superior overall performance in a scenario. Evaluations are implemented at epochs 10, 25, and 50. Fig.~\ref{fig:ab_data_exp} illustrates several key findings:
\begin{table}
\centering
\caption{
Data sampling proportions across the three training stages (early, middle, and final). Ep. denotes the number of training epochs per stage. Values indicate Sound/Speech/Speech-and-Sound sampling weights.
}
\label{tab:ab_data_schedules_detailed}
\small
\begin{tabular}{@{}l ccc@{}}
\toprule
\textbf{Strategy} & Early (Ep.\ 0-10) & Middle (Ep.\ 10-25) & Final (Ep.\ 25-50) \\ \midrule
Constant & 0.33/0.33/0.33 & 0.33/0.33/0.33 & 0.33/0.33/0.33 \\
Gradual  & 0.40/0.40/0.20 & 0.40/0.20/0.40 & 0.25/0.25/0.50 \\
Decoupled & 0.50/0.50/0.00 & 0.50/0.50/0.00 & 0.00/0.00/1.00 \\ \bottomrule
\end{tabular}
\end{table} 
(1) \textbf{Robustness of Constant Balancing.} \textit{Constant} generally outperforms other strategies, suggesting that a uniform data curriculum is vital for robust unified generation without favoring specific categories at the expense of 
others. This also demonstrates that our framework does not require meticulous ratio tuning for convergence. 
(2) \textbf{Catastrophic Forgetting in Decoupled Sequential Training.} In \textit{Gradual} and \textit{Decoupled}, sound and speech performance degrades as speech-and-sound data increases. This is most acute in \textit{Decoupled}, where the abrupt shift to exclusive speech-and-sound training triggers significant forgetting, emphasizing the necessity of continuous multi-scenario exposure. (3) \textbf{Synergistic Transfer from Speech-and-Sound.} \textit{Constant} and \textit{Gradual} outperform \textit{Decoupled} in sound/speech even when \textit{Decoupled} focuses solely on those scenarios. This reveals a positive transfer effect: mastering speech-and-sound scenarios, which requires decoupling complex acoustic components in prompts, forces the model to learn more discriminative features, thereby accelerating proficiency in single-scenario generation.

\subsection{Impact of Data Augmentation} \label{app:ablation_data_enhance} 

To investigate the impact of annotation diversity, we compare two training settings: 1) \textit{Full Annotation}: All 5 annotations per clip. 2) \textit{Single Annotation}: One annotation per clip.  Both variants follow the main experimental setup, but are trained for 25 epochs on a one-third data subset. As shown in Tab~\ref{tab:app_data_enhance}, \textit{Full Annotation} consistently achieves superior semantic alignment and acoustic quality. This suggests that exposing the model to varied phrasing and multiple annotations enhances generalization to Free-Form prompts and mitigates overfitting to specific scenarios.

\begin{table}[H]
\centering
\caption{Impact of annotation diversity. Comparison between using five text annotations (Full) versus one (Single) per clip throughout training. Full annotation consistently yields superior performance.}
\label{tab:app_data_enhance}
\resizebox{\linewidth}{!}{\begin{tabular}{lcccccccccc} 
\toprule
\multicolumn{1}{c}{} & \multicolumn{5}{c}{Composite} & \multicolumn{3}{c}{Sound} & \multicolumn{2}{c}{Speech} \\ \cmidrule(lr){2-6} \cmidrule(lr){7-9} \cmidrule(lr){10-11}
\multicolumn{1}{c}{\multirow{-2}{*}{Setting}} & \multicolumn{1}{c}{FD $\downarrow$} & \multicolumn{1}{c}{KL $\downarrow$} & \multicolumn{1}{l}{CLAP $\uparrow$} & \multicolumn{1}{c}{WER $\downarrow$} & \multicolumn{1}{c}{UTMOS $\uparrow$} & \multicolumn{1}{l}{FD $\downarrow$} & \multicolumn{1}{l}{KL $\downarrow$} & \multicolumn{1}{l}{CLAP $\uparrow$} & \multicolumn{1}{c}{WER $\downarrow$} & \multicolumn{1}{c}{UTMOS $\uparrow$} \\  \midrule
\rowcolor[HTML]{CFE2F3}[7pt][7pt]  Full Annotation & 234.89 & 1.07 & 0.17 & 0.81 & 2.21 & 	251.87 & 1.94 & 0.16 & 	0.36 & 	3.03 \\
Single Annotation & 245.13 & 1.07 & 0.16 & 0.85 &  2.21 & 273.66 & 2.06 & 0.12 & 0.44 & 2.98 \\
 \bottomrule
\end{tabular}}
\end{table}

\subsubsection{Annotation Quality of Speech-and-Sound Subset in \benchname}
\label{app:data_annotation_quality}
Since speech-and-sound data is annotated automatically, we verify the reliability through human inspection. Three annotators independently rated a random subset of 91 items, yielding 156 judgments in total. Each judgment is binary and assesses whether the prompt faithfully describes the paired audio along three dimensions: the accuracy of the speech transcription, the accuracy of the non-linguistic description, and the correctness of the stated temporal relations. Temporal relations are judged as categorical event ordering rather than millisecond-level boundary alignment, matching the granularity at which the prompts are written. Scores are averaged over all judgments, and an item counts as overall correct only when all three dimensions are satisfied.

\begin{table}[h]
\centering
\caption{Human verification of Speech-and-Sound annotations in \benchname (156 judgments on 91 randomly selected items).}
\label{tab:annotation_quality}
\begin{tabular}{lc}
\toprule
Dimension & Accuracy $\uparrow$ \\
\midrule
Speech Transcription & 98.1\% \\
Sound Description & 99.4\% \\
Temporal Relation & 100.0\% \\
\midrule
Overall & 96.8\% \\
\bottomrule
\end{tabular}
\end{table}

Accuracy is high across all three dimensions, indicating that the decoupled annotation pipeline produces prompts that can serve as reliable references for joint audio evaluation.

\subsubsection{Interaction Type Coverage of Speech-and-Sound Subset in \benchname} \label{app:data_interaction_types}

To characterize what kinds of speech-and-sound interaction the benchmark covers, we label all 4,500 speech-and-sound samples with a five-way taxonomy using Gemini-2.5 Pro. Each sample receives a single label under a fixed priority rubric, so samples exhibiting more than one interaction are assigned to the highest-priority category rather than counted multiple times.

\begin{table}[h]
\centering
\caption{Distribution of interaction types across the 4,500 speech-and-sound subset of the \benchname. Each sample is assigned a single label under a fixed priority rubric.}
\label{tab:interaction_types}
\begin{tabular}{lcc}
\toprule
Interaction Type & Count & Ratio (\%) \\
\midrule
Sequential ordering & 605 & 13.4 \\
Overlapping speech--sound & 698 & 15.5 \\
Background ducking / loudness adaptation & 739 & 16.4 \\
Causal trigger & 1,198 & 26.6 \\
Rhythmic alignment & 1,260 & 28.0 \\
\midrule
Total & 4,500 & 100.0 \\
\bottomrule
\end{tabular}
\end{table}

The distribution is broad, with no category falling below $13\%$, so the benchmark does not reduce to a single dominant interaction pattern. Within the sequential ordering category, the number of ordered events ranges from two to six, with two-event cases accounting for $49.9\%$; the remainder therefore requires the model to maintain longer temporal chains.

\subsection{Ablation Study}
Full-scale training is costly, so all ablation experiments below are conducted under a common reduced-scale setting: 15 epochs on a 100k subset with an average data sampling proportion unless stated otherwise. Since this budget is considerably smaller than that of the main experiments, the absolute values are therefore not comparable to the main results;  only the relative ordering is meaningful.

\subsubsection{Planning Horizon} \label{app:ablation_planning_horizon}
The planning horizon $K$ determines how many latent anchors the model emits before entering acoustic generation. To justify our choice, we compare $K \in \{3, 6, 12\}$ under an identical setup. Results are shown in Tab.~\ref{tab:ablation_planning_horizon}.

\begin{table}[h]
\centering
\caption{Ablation on the planning horizon $K$. All variants are trained for 15 epochs on a 100k subset. KL is computed via the PaSST classifier.}
\label{tab:ablation_planning_horizon}
\begin{tabular}{lcccc}
\toprule
& \multicolumn{2}{c}{Speech-and-Sound} & Sound & Speech \\
\cmidrule(lr){2-3} \cmidrule(lr){4-4} \cmidrule(lr){5-5}
Variant & KL $\downarrow$ & WER $\downarrow$ & KL $\downarrow$ & WER $\downarrow$ \\
\midrule
$K=3$ & 1.15 & 1.21 & 2.49 & 1.11 \\
$K=6$ (PlanAudio) & 1.12 & 1.14 & 2.17 & 1.10 \\
$K=12$ & 1.10 & 1.14 & 2.22 & 1.08 \\
\bottomrule
\end{tabular}
\end{table}

Increasing $K$ from 3 to 6 improves speech-and-sound intelligibility (WER $1.21 \to 1.14$) and sound fidelity (KL $2.49 \to 2.17$), indicating that three anchors are too coarse to represent the event structure of a speech-and-sound scene. Enlarging the horizon further to $K=12$ yields no consistent gain while doubling the length of the planning phase, so we adopt $K=6$ as the setting that attains the same quality at lower inference cost.

\subsubsection{Fixed Planning Horizon and Variable-Length Audio}
\label{app:ablation_variable_length}
A fixed $K$ constrains only the number of semantic anchors, not the duration of the generated audio. The two phases operate at different granularities: AF3Encoder features are extracted chronologically and pooled into $K$ anchors that form a coarse temporal backbone, while the acoustic phase renders frame-level prosodic contours and event boundaries conditioned on that backbone. Audio length is decided in this second phase, which terminates on the \texttt{<|eoa|>} token, so prompts describing longer or denser scenes are decoded to correspondingly longer sequences without truncation.

We also examined whether the horizon should itself be predicted. Prior work on continuous latent reasoning in text LLMs reports that fixed- and variable-length chains perform comparably~\cite{latent_cot_llm2}, and our findings agree in the audio setting: a variant that terminates the planning phase through a binary classifier underperformed the fixed-length configuration, as length prediction proved unstable to optimize and tended to terminate prematurely under dense acoustic mixtures. We therefore retain a fixed horizon throughout.

\subsubsection{Sensitivity to the Loss Weight}
\label{app:ablation_loss_weight}

We vary the weight $\lambda_1$ of the semantic latent regression loss while keeping the acoustic objective fixed at $\lambda_2 = 1.0$. Results are shown in Tab.~\ref{tab:ablation_loss_weight}.

\begin{table}[h]
\centering
\caption{Sensitivity to the semantic latent loss weight $\lambda_1$, with the acoustic objective fixed at $\lambda_2=1.0$. KL is computed via the PaSST classifier.}
\label{tab:ablation_loss_weight}
\begin{tabular}{lcccc}
\toprule
& \multicolumn{2}{c}{Composite} & Sound & Speech \\
\cmidrule(lr){2-3} \cmidrule(lr){4-4} \cmidrule(lr){5-5}
Variant & KL $\downarrow$ & WER $\downarrow$ & KL $\downarrow$ & WER $\downarrow$ \\
\midrule
$\lambda_1 = 0.2$ & 1.14 & 1.19 & 2.62 & 1.11 \\
$\lambda_1 = 1.0$ (PlanAudio) & 1.12 & 1.14 & 2.17 & 1.10 \\
$\lambda_1 = 5.0$ & 1.16 & 1.13 & 2.08 & 1.08 \\
\bottomrule
\end{tabular}
\end{table}

Weakening the semantic objective to $\lambda_1 = 0.2$ degrades sound generation appreciably (KL $2.62$ vs.\ $2.17$), confirming that the planning phase requires sufficient supervision to be useful. Beyond this point the model is largely insensitive to the setting: $\lambda_1 = 1.0$ and $\lambda_1 = 5.0$ differ only marginally and trade places across metrics, so PlanAudio does not depend on careful tuning of this hyperparameter. We adopt $\lambda_1 = 1.0$, which balances the two objectives without scaling up gradient magnitudes in the planning phase.

\subsection{Multi-stream Token Modeling with Delay Pattern} \label{app:delay}

In the context of multi-stream audio token modeling, the delay pattern~\cite{musicgen} is a sequence interleaving strategy designed to handle parallel audio tokens from a multi-codebook audio codec model. Unlike the flattening pattern, which serializes all tokens into a long 1D sequence, the delay pattern maintains the original temporal resolution by introducing a fixed offset between codebook streams.
\begin{figure}[H]
    \centering
    \includegraphics[width=0.7\linewidth]{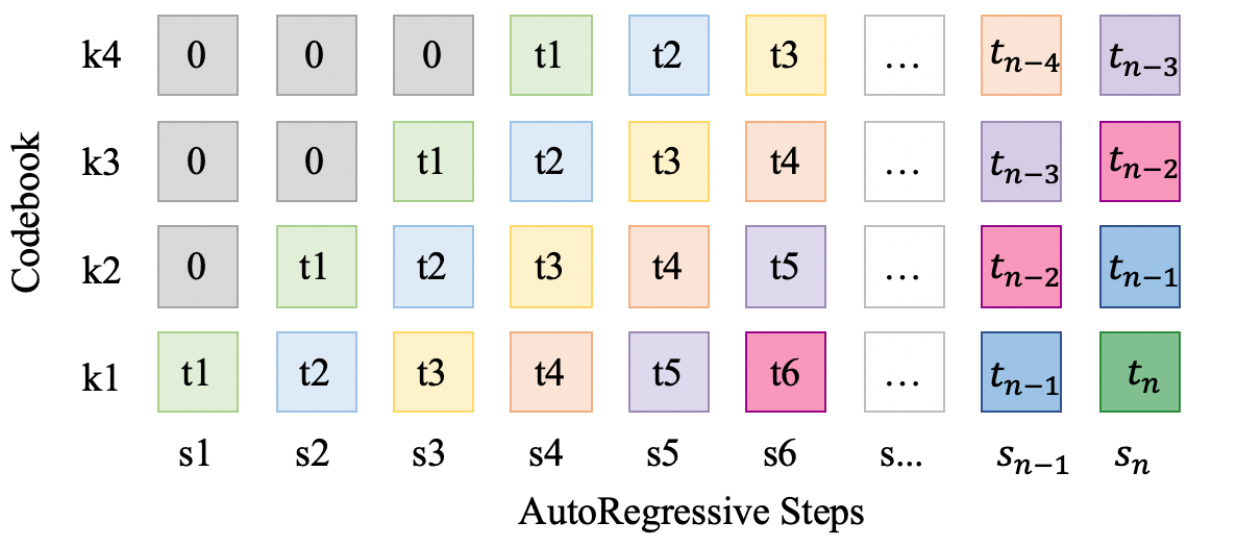}
    \caption{Delay sequence interleaving pattern. Each time step $t_1$, $t_2$, ..., $t_n$ is composed of 4 quantized values (corresponding to $k_1$, ..., $k_4$). When doing autoregressive modelling, we interleave them and result in a new sequence with 4 parallel streams and steps $s_1$, $s_2$, ..., $s_n$. 0 indicates empty positions in the pattern.}
    \label{app:fig_delay}
\end{figure}
As shown in Figure~\ref{app:fig_delay}, for a codec with $K$ codebooks, the $k$-th stream is shifted to the right by $k-1$ time steps. This data arrangement allows a single autoregressive Transformer to predict $K$ tokens at each logical step through parallel linear output heads. In this configuration, the prediction of the $k$-th codebook at time $t$ is conditioned on the previous $t-1$ steps of all streams, as well as the concurrent tokens of the $k-1$ higher-level codebooks due to the temporal shift. This mechanism effectively transforms the parallel codebook structure into a format compatible with autoregressive modeling without increasing the total number of transformer inference steps.

\subsection{Balancing Reconstruction Quality and Generative Performance across Audio Domains} \label{app:codec_exp}
In this section, we present an experimental analysis of PlanAudio using different audio codecs. Since our task encompasses speech, sound, and composite scenarios, it is essential to select a universal codec model. Given that the reconstruction quality of a codec determines the performance ceiling of the generative model, we first evaluated the reconstruction quality of candidate codecs across various audio categories. The reconstruction results are summarized in Table~\ref{app:tab_recons}.

Regarding reconstruction performance, among multi-layer codecs, AudioCraft and DAC exhibit lower frame rates alongside higher reconstruction quality, differing primarily in codebook size and quantity. Among single-layer codecs, UniCodec achieves an optimal balance between sound and speech reconstruction. Consequently, we conducted PlanAudio training experiments across these three codecs while keeping all other parameters identical. It is worth noting that due to the parallel delay pattern employed for multi-stream audio token prediction, there was no significant difference in training time per epoch among these configurations. The results of PlanAudio on various codecs are presented in Table~\ref{app:tab_codec_exp}.

\begin{table}[H]
\centering
\caption{Comparison of reconstruction quality across various audio codecs and scenarios. The results demonstrate the performance ceiling of candidate codecs across sound, speech, and composite scenarios.}
\label{app:tab_recons}
\resizebox{\linewidth}{!}{
\begin{tabular}{lccccccccccccc}
\toprule
\multicolumn{1}{l}{\multirow{2}{*}{Codec}} & \multicolumn{1}{c}{\multirow{2}{*}{Nq}} & \multicolumn{1}{c}{\multirow{2}{*}{\begin{tabular}[c]{@{}c@{}}Codebook\\ Size\end{tabular}}} & \multicolumn{1}{c}{\multirow{2}{*}{\begin{tabular}[c]{@{}c@{}}Frame\\ Rate\end{tabular}}} & \multicolumn{5}{c}{Composite} & \multicolumn{3}{c}{Sound} & \multicolumn{2}{c}{Speech} \\ \cmidrule(lr){5-9} \cmidrule(lr){10-12} \cmidrule(lr){13-14} 
\multicolumn{1}{c}{} & \multicolumn{1}{c}{} & \multicolumn{1}{c}{} & \multicolumn{1}{c}{} & \multicolumn{1}{c}{FD $\downarrow$} & \multicolumn{1}{c}{KL $\downarrow$} & \multicolumn{1}{c}{CLAP $\uparrow$} & \multicolumn{1}{c}{WER $\downarrow$} & \multicolumn{1}{c}{UTMOS $\uparrow$} & \multicolumn{1}{c}{FD $\downarrow$} & \multicolumn{1}{c}{KL $\downarrow$} & \multicolumn{1}{c}{CLAP $\uparrow$} & \multicolumn{1}{c}{WER $\downarrow$} & \multicolumn{1}{c}{UTMOS $\uparrow$} \\ \midrule
GT & \multicolumn{1}{c}{-} & - & - & 0.00 & 0.00 & 0.17 & 0.10 & 2.69 & 0.29 & 0 & 0.29 & \multicolumn{1}{c}{-} & \multicolumn{1}{c}{-} \\ \midrule
\multicolumn{3}{l}{Single-Layer Codec} & \multicolumn{1}{c}{} & \multicolumn{1}{c}{} & \multicolumn{1}{c}{} & \multicolumn{1}{c}{} & \multicolumn{1}{c}{} & \multicolumn{1}{c}{} & \multicolumn{1}{c}{} & \multicolumn{1}{c}{} & \multicolumn{1}{c}{} & \multicolumn{1}{c}{} & \multicolumn{1}{c}{} \\ \midrule
Xcodec & 1 & 1024 & 50 & \textbf{217.31} & \textbf{0.59} & 0.19 & 0.46 & 2.56 & \textbf{322.20} &  \underline{1.48} &  \underline{0.10} & 0.11 & 3.26 \\
Unicodec & 1 & 8192 & 75 & \underline{370.24} & \underline{0.65} & \textbf{0.23} & \textbf{0.27} & \textbf{2.60} & {\ul 408.13} & \textbf{1.31} & 0.16 & \textbf{0.04} & \underline{3.57} \\
Wav\_Tokenizer & 1 & 4096 & 75 & 533.90 & 1.01 & \underline{0.20} & \underline{0.35} & 2.53 & 608.10 & 2.24 & 0.06 & \underline{0.05} & \textbf{3.67} \\ \midrule
\multicolumn{3}{l}{Multi-Layer Codec} & \multicolumn{1}{c}{} & \multicolumn{1}{c}{} & \multicolumn{1}{c}{} & \multicolumn{1}{c}{} & \multicolumn{1}{c}{} & \multicolumn{1}{c}{} & \multicolumn{1}{c}{} & \multicolumn{1}{c}{} & \multicolumn{1}{c}{} & \multicolumn{1}{c}{} & \multicolumn{1}{c}{} \\ \midrule
Audiocraft & 4 & 2048 & 50 & \underline{112} & 0.37 & 0.21 & 0.21 & 2.43 & 130 & 0.48 & 0.23 & 0.04 & 3.13 \\
Encodec & 8 & 1024 & 75 & 198 & 0.40 & \underline{0.22} & 0.19 & 2.45 & 252 & 0.47 & 0.32 & 0.03 & 3.45 \\
Xcodec & 8 & 1024 & 50 & 155 & \underline{0.34} & \textbf{0.34} &  & \multicolumn{1}{l}{} & 285 & 0.95 & 0.18 & \multicolumn{1}{c}{} & \multicolumn{1}{c}{} \\
DAC & 12 & 1024 & 50 & \textbf{83.8} & \textbf{0.23} & \underline{0.22} & \textbf{0.12} & \textbf{2.65} & \textbf{96.8} & \textbf{0.29} & 0.25 & 0.03 & \textbf{3.58} \\
\bottomrule
\end{tabular}}
\end{table}

\begin{table}[H]
\centering
\caption{Generative performance of PlanAudio across different audio codecs and planning configurations. The results are evaluated across composite, sound, and speech scenarios.}
\label{app:tab_codec_exp}
\resizebox{\linewidth}{!}{\begin{tabular}{lcccccccccc}
\toprule
\multicolumn{1}{c}{} & \multicolumn{5}{c}{Composite} & \multicolumn{3}{c}{Sound} & \multicolumn{2}{c}{Speech} \\ \cmidrule(lr){2-6} \cmidrule(lr){7-9} \cmidrule(lr){10-11} 
\multicolumn{1}{c}{\multirow{-2}{*}{model}} & \multicolumn{1}{l}{FD $\downarrow$} & \multicolumn{1}{l}{KL $\downarrow$} & \multicolumn{1}{l}{CLAP $\uparrow$} & \multicolumn{1}{c}{WER $\downarrow$} & \multicolumn{1}{c}{UTMOS $\uparrow$} & \multicolumn{1}{l}{FD $\downarrow$} & \multicolumn{1}{l}{KL $\downarrow$} & \multicolumn{1}{l}{CLAP $\uparrow$} & \multicolumn{1}{l}{WER $\downarrow$} & \multicolumn{1}{l}{UTMOS $\uparrow$} \\ \midrule
\rowcolor[HTML]{CFE2F3} 
DAC-Planning & 429 & 1.64 & 0.19 & 0.81 & 2.72 & 583 & 3.78 & 0.00 & 0.16 & 3.40 \\
DAC-w/o Planning & 484 & 2.37 & 0.13 & 0.84 & 2.11 & 622 & 4.78 & -0.05 & 0.40 & 3.23 \\ \midrule
\rowcolor[HTML]{CFE2F3} 
UniCodec-Planning & 415 & 1.11 & 0.23 & 0.35 & 2.81 & 477 & 1.87 & 0.20 & 0.06 & 3.61 \\
UniCodec-w/o Planning & 472 & 1.37 & 0.22 & 0.43 & 2.59 & 494 & 2.83 & 0.08 & 0.08 & 3.59 \\ \midrule
\rowcolor[HTML]{CFE2F3} 
AudioCraft-Planning & 180 & 1.05 & 0.21 & 0.47 & 2.49 & 233 & 1.89 & 0.19 & 0.11 & 3.11 \\
AudioCraft-w/o Planning & 252 & 1.38 & 0.18 & 0.57 & 2.35 & 281 & 3.10 & 0.04 & 0.21 & 3.04 \\ \bottomrule
\end{tabular}}
\end{table}

From the perspective of Plan, the Planning mechanism consistently yields significant improvements over the version w/o Plan for every codec, demonstrating the robustness of our approach. From the codec perspective, although DAC achieves the best reconstruction quality, it performs the worst in terms of actual generation. We attribute this phenomenon primarily to its excessive number of codebooks, which makes it difficult for the model to learn the underlying patterns. For UniCodec and AudioCraft, by comparing the generation performance in Table~\ref{app:tab_codec_exp} with the reconstruction performance in Table~\ref{app:tab_recons}, we observe a strong correlation between the two. UniCodec, which performs better in speech reconstruction, also excels in speech generation. Similarly, AudioCraft, which is adept at sound reconstruction, shows consistent performance during generation. 

Given the current results, we have yet to find a single codec that perfectly balances reconstruction quality and generation performance across all audio domains; thus, a trade-off is inevitable. Ultimately, we selected AudioCraft as the discrete audio tokenizer for PlanAudio.

\subsection{The Semantic Coverage Factor}
\label{app:eval_scf}

The Semantic Coverage Factor (SCF) quantifies semantic omission, i.e.\ the extent to which events named in the prompt are actually realized in the generated audio. We first filter out low-probability background noise from the generated audio, then pair each detected event with the events named in the prompt. Every pair contributes the cosine similarity of the two, weighted by the model's confidence in the detected event, and only pairs whose similarity exceeds a threshold $t$ are counted. The accumulated score is normalized by the number of ground-truth events, so a value of $1.0$ indicates that every prompted event is realized with high confidence.

\subsection{Details about Human Study} \label{app:eval_user}
In our human study, each of the 50 samples was rated by 10 raters across four models and four criteria, yielding 8,000 observations in total. We provide the instructions for the subjective evaluation:

\begin{center}
    \label{app:tab_user}
    \vspace{-3pt} 
    \begin{closedprompt}
    This evaluation contains 50 samples in total. Each subfolder includes audio generated by four different models from the same text prompt, along with a textual description provided in description.txt. The correspondence between model names and audio files has been randomly shuffled.

        Please listen to the audio samples and read the text description, then rate each sample according to the following four dimensions.
        
        Scoring Criteria:
        \begin{itemize}
            \item Audio Quality: Score range: 0–5, with 0.5-point increments. Higher scores indicate better audio quality.
            \item Temporal Correctness: Evaluate whether the ordering of audio events matches the temporal sequence described in the text. Higher consistency receives higher scores. Pay particular attention to temporal cues such as “after”, “followed by”, etc.
            \item Semantic Alignment: Score range: 0–5, with 0.5-point increments. Evaluate how completely the events mentioned in the text are reflected in the audio. More comprehensive semantic coverage receives higher scores. For example, if the text mentions bird chirping and human speech, both should be present in the generated audio.
            \item Audio Authenticity: Score range: 0–5, with 0.5-point increments. Evaluate whether the environmental sounds and speech are naturally integrated, rather than sounding like a simple overlap of two separate audio tracks. Higher scores indicate more realistic and coherent audio.
        \end{itemize}
    Note: When rating a specific dimension, please focus only on the corresponding criterion without considering other evaluation dimensions.
    
    \end{closedprompt}
\end{center}